\documentclass{article}
\usepackage{spconf,amsmath,graphicx}
\usepackage{xcolor}
\usepackage{lipsum}
\usepackage{amssymb}
\usepackage{tabularx} 
\usepackage{dblfloatfix}
\usepackage{placeins}
\usepackage{comment}
\usepackage[ruled,vlined]{algorithm2e}
\usepackage{algpseudocode}

\usepackage{float}
\usepackage{enumitem}
\usepackage{mathrsfs}
\usepackage{mathtools}
\usepackage{ctable}
\usepackage{balance}

\title{Streamlined Hybrid Annotation Framework using Scalable\\ Codestream for Bandwidth-Restricted UAV Object Detection}

\name{Karim El Khoury\textsuperscript{$\star$}, Tiffanie Godelaine\textsuperscript{$\star$}, Simon Delvaux, Sébastien Lugan, and Benoît Macq\thanks{\textsuperscript{$\star$}The authors have contributed equally to this work.}}

\address{ICTEAM, Université Catholique de Louvain, Louvain-la-Neuve, Belgium}

\begin{document}

\maketitle

\begin{abstract}
Emergency response missions depend on the fast relay of visual information, a task to which unmanned aerial vehicles are well adapted. However, the effective use of unmanned aerial vehicles is often compromised by bandwidth limitations that impede fast data transmission, thereby delaying the quick decision-making necessary in emergency situations. To address these challenges, this paper presents a streamlined hybrid annotation framework that utilizes the JPEG 2000 compression algorithm to facilitate object detection under limited bandwidth. The proposed framework employs a fine-tuned deep learning network for initial image annotation at lower resolutions and uses JPEG 2000's scalable codestream to selectively enhance the image resolution in critical areas that require human expert annotation. We show that our proposed hybrid framework reduces the response time by a factor of 34 in emergency situations compared to a baseline approach.
\end{abstract}


\section{Introduction}
\label{sec:intro}

The utilization of Unmanned Aerial Vehicles (UAVs) has become standard practice in improving the operational capabilities of first responders in emergency situations. UAVs provide essential real-time visual information for areas affected by disasters, aiding in damage assessment, hazard detection, and search and rescue missions \cite{Papić_Šolić_Milan_Gotovac_Polić_2021,Tuśnio_Wróblewski_2021}. 
Hybrid annotation stands out as a method to improve the accuracy and reliability of data interpretation in such challenging scenarios. Hybrid annotation begins with an automated Deep Learning (DL)-based object detection, which is then refined by human annotators in areas of uncertainty. The aim is to enhance the annotation process to accomplish two essential objectives: (1) allowing first responders to make fast and accurate decisions, and (2) saving human expert annotators’ valuable time. However, the effectiveness of conventional hybrid annotation frameworks depends on the reliability of data transmission. In emergency scenarios, limited bandwidth, obstructed line of sight and damaged infrastructure delay UAV-to-ground data transmission, hindering the decision-making process.

In this paper, we put forward a streamlined hybrid annotation framework using scalable codestream for bandwidth-restricted UAV object detection. Our framework incorporates the JPEG 2000 (J2K) compression algorithm within a hybrid annotation workflow. The scalability of the J2K codestream allows for multiresolution access to specific regions of interest, particularly useful when dealing with limited bandwidth \cite{Skodras_Christopoulos_Ebrahimi_2001}. The proposed approach streamlines the hybrid annotation process by initially analyzing low-resolution images through a fine-tuned DL network. For areas requiring more detailed analysis, the J2K algorithm facilitates selective enhancement at high resolution, enabling human experts to perform precise annotations.

\vspace{1mm}

\noindent The contributions of our work are summarized as follows:
\begin{itemize}[leftmargin=*]

 \item We propose an optimized hybrid annotation framework that takes into account the scenario-specific bandwidth constraints to ensure a fast decision-making process. We observe that the proposed framework reduces the response time by a factor of 34 compared to a baseline approach. 

 \item We show that fine-tuning a DL network for object detection on low-resolution images remains effective while providing fast information access in bandwidth-restricted situations.

\end{itemize}

\section{Related Works}
\label{sec:relatedworks}

In order to categorize the contribution of each of the related works, we define the following four research challenges: (I) High-resolution image compression, (II) DL-based object detection, (III) Hybrid annotation, and (IV) Applicability to bandwidth-restricted UAV scenarios. Table \ref{tab:related_work} provides a summarized comparison of the related works as per the aforementioned challenges. We focus on contributions that address at least two of the four challenges. To the best of our knowledge, the proposed streamlined hybrid annotation framework using scalable codestream for bandwidth-restricted UAV object detection is the first contribution to offer a comprehensive solution to all four challenges.

\vspace{-3mm}

\subsection{Related works merging (I) and (IV):}

Zhou et al. \cite{Zhou_Deng_Zhao_Xia_Li_Chen_2015} present a comprehensive review of remote sensing image compression algorithms. They highlight the J2K codec's ability to achieve high compression ratios without significant loss of image quality which is crucial for efficient transmission of high-resolution images in limited bandwidth scenarios. Indradjad et al. \cite{Indradjad_Nasution_Gunawan_Widipaminto_2019} compare four wavelet-based methods for compressing satellite imagery. They show that J2K delivers better image quality than other wavelet-based codecs at lower bit rates.

\vspace{-3mm}

\subsection{Related works merging (I) and (II):}

Ehrlich et al. \cite{Ehrlich_Davis_Lim_Shrivastava_2021} introduce a self-supervised artifact correction approach that improves object detection accuracy on JPEG compressed images. Ghazvinian Zanjani et al. \cite{GhazvinianZanjani_Zinger_Piepers_Mahmoudpour_Schelkens_2019} establish that DL models can accurately identify metastatic cancer in J2K compressed histopathological images. El Khoury et al. \cite{ElKhoury_Fockedey_Brion_Macq_2021,ElKhoury_Wynen_Maggi_BachCuadra_Macq_2024} show that fine-tuning a 3D U-Net improves its ability to handle J2K compression artifacts while maintaining precise organ detection and segmentation on radiotherapy scans.

\vspace{-3mm}

\subsection{Related works merging (I), (II) and (IV):}

Zhang et al. \cite{Zhang_Weng_Luo_Liu_Yang_Lin_Zhang_Li_2018} develop a DL-based mechanism for UAV-aided networks that compresses media data into constant-sized latent codes for efficient real-time transmission, successfully maintaining the quality of visual data for small objects despite bandwidth constraints and compression artifacts. However, their proposed system relies on having uninterrupted line-of-sight communication which limits its applicability in obstructed and complex environments. Preethy Byju et al. \cite{Preethy_Sumbul_Demir_Bruzzone_2021} design a DL framework for efficiently classifying scenes in J2K compressed remote-sensing images, which balances computational efficiency with detection accuracy. Given that they work on UAV scenarios with large compressed data repositories, their contribution is not tailored for real-time use. Both contributions do not work within a hybrid annotation framework.

\vspace{-3mm}

\subsection{Related works merging (II), (III) and (IV):}

Kellenberger et al. \cite{Kellenberger_Marcos_Lobry_Tuia_2019} present an active learning method that requires limited labeled data from human annotators to improve animal detection in UAV imagery. 
Yamani et al. \cite{Yamani_Alyami_Luqman_2024} introduce a single-stage object detection active learning framework for UAV imagery that minimizes annotation expenses through an innovative uncertainty aggregation technique while mitigating class imbalance. Both contributions do not consider the image compression requirements imposed by low-bandwidth constraints.

\begin{table}[!t]
\centering
\small
\caption{Comparison of related works with respect to four research challenges: (I) High-resolution image compression, (II) DL-based object detection, (III) Hybrid annotation, and (IV) Applicability to bandwidth-restricted UAV scenarios.\\}
\label{tab:related_work}
\setlength\tabcolsep{1pt} 
\begin{tabularx}{\columnwidth}{*{12}{>{\centering\arraybackslash}X}}
\hline
  & \textbf{\cite{Zhou_Deng_Zhao_Xia_Li_Chen_2015}} & \textbf{\cite{Indradjad_Nasution_Gunawan_Widipaminto_2019}} & \textbf{\cite{Ehrlich_Davis_Lim_Shrivastava_2021}} & \textbf{\cite{GhazvinianZanjani_Zinger_Piepers_Mahmoudpour_Schelkens_2019}} & \textbf{\cite{ElKhoury_Fockedey_Brion_Macq_2021}} & \textbf{\cite{ElKhoury_Wynen_Maggi_BachCuadra_Macq_2024}} & \textbf{\cite{Zhang_Weng_Luo_Liu_Yang_Lin_Zhang_Li_2018}} & \textbf{\cite{Preethy_Sumbul_Demir_Bruzzone_2021}} & \textbf{\cite{Kellenberger_Marcos_Lobry_Tuia_2019}} & \textbf{\cite{Yamani_Alyami_Luqman_2024}}  & \textbf{Ours} \\ \hline
\textbf{(I)}   & \checkmark & \checkmark & \checkmark & \checkmark & \checkmark & \checkmark & \checkmark & \checkmark & \          &            & \checkmark \\
\textbf{(II)}  &            &            & \checkmark & \checkmark & \checkmark & \checkmark & \checkmark & \checkmark & \checkmark & \checkmark & \checkmark \\
\textbf{(III)} &            &            &            &            &            &            &            &            & \checkmark & \checkmark & \checkmark \\
\textbf{(IV)}  & \checkmark & \checkmark &            &            &            &            & \checkmark & \checkmark & \checkmark & \checkmark & \checkmark \\  \hline
\end{tabularx}
\end{table}

\section{Proposed Approach}
\label{sec:approach}

In this section, we first present a conventional baseline hybrid annotation framework and then detail our proposed streamlined hybrid annotation framework. It should be noted that both frameworks are based on the J2K scalable codestream  for encoding and decoding as well as multiresolution tile extraction.

\begin{figure*}[!t]
  \centering
  \includegraphics[width=0.91\linewidth]{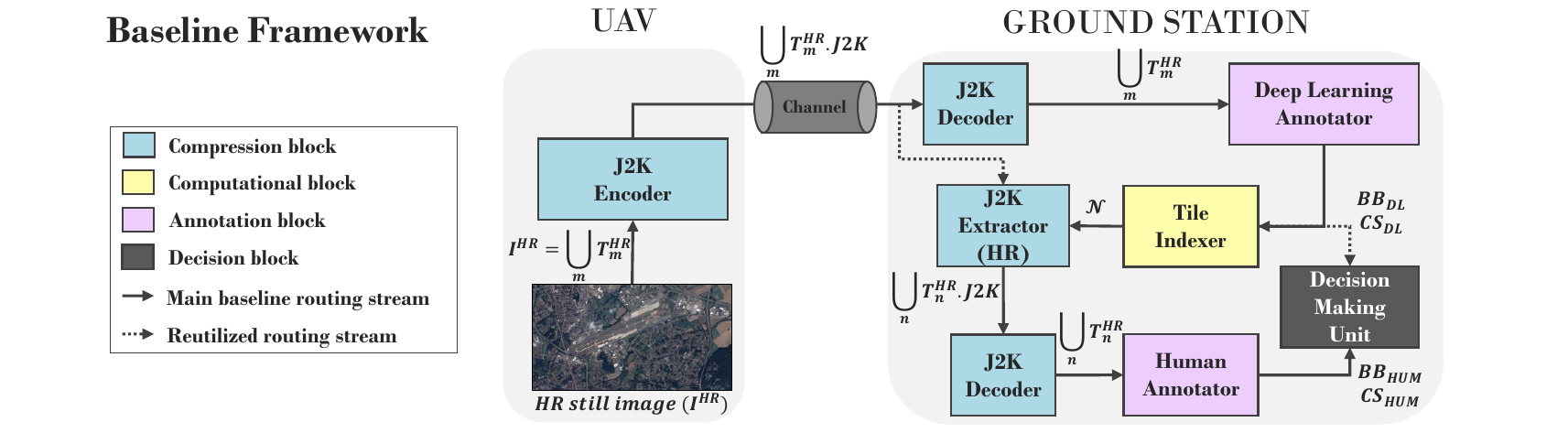} 
  \caption{Simplified block diagram of the baseline hybrid annotation framework.}
  \label{fig:baseline}
\end{figure*}

\vspace{-0.5mm}

\subsection{Baseline hybrid annotation framework}

The baseline framework aims to emulate the standard hybrid annotation approach used by UAVs in emergency situations. Figure \ref{fig:baseline} illustrates a simplified block diagram of the framework, modeling a conventional compression scheme through a lossless UAV-to-ground station communication
channel. A step-by-step pseudo-code of the baseline hybrid annotation framework is presented in Algorithm 1.

\SetKwFunction{FSize}{SIZE}

The basic working principles of the baseline framework are described as follows. The framework takes as input a high-resolution still image ($I^{HR}$), a predefined high-resolution level ($HR$), the high-resolution tile size for the J2K encoding (\FSize{$T^{HR}$}), and a predetermined human annotation budget ($|\mathscr{N}|$), which represents how many tiles a human annotator can afford to annotate in time-limited situations. The first step involves calculating the total number of tiles ($|\mathscr{M}|$) in $I^{HR}$. The high-resolution tiles ($T^{HR}$) from $I^{HR}$ are then encoded using J2K and sent via a lossless communication channel to the ground station. The tiles are then decoded and processed by a DL model to produce preliminary annotations. The \textit{indexer} block, detailed in Subroutine 1, processes the resultant annotations, made up of the bounding boxes ($BB_{DL}$) and their confidence scores ($CS_{DL}$). It identifies a set of tiles that require inspection by a human expert annotator and stores their indices ($\mathscr{N}$). The framework then calls the \textit{extractor} block, detailed in Subroutine 2, to process the J2K scalable codestream and extracts the specific $\mathscr{N}$ tiles in high resolution. The $\mathscr{N}$ selected tiles ($T^{HR}_n$) are then annotated by the human expert. The resultant bounding boxes ($BB_{HUM}$) and confidence scores ($CS_{HUM}$) are sent out along with the DL annotations ($BB_{DL}$, $CS_{DL}$) to the decision-making unit where critical decisions are taken to respond to emergency situations.

Essentially, the main limitation of the baseline framework is due to its constant transmission of high-resolution tiles from the UAV to the ground station, regardless of the scenario-specific constraints. This becomes problematic when data rates and time demands are low, thereby impeding the framework's effectiveness in time-sensitive situations.

\newpage

\begin{algorithm}[!h]
\small
\caption{Baseline framework}
\DontPrintSemicolon
\SetKwProg{SFn}{Function}{:}{}
\SetKwFunction{Fmain}{main}
\SetKwFunction{Ftile}{BC}
\SetKwFunction{FJcomp}{J2K}
\SetKwFunction{FExt}{EXT}
\SetKwFunction{FJdecomp}{J2K$^{-1}$}
\SetKwFunction{FDL}{DL}
\SetKwFunction{FIndex}{IND}
\SetKwFunction{FHUM}{HUM}
\SetKwFunction{FTR}{TR}
\SetKwFunction{FSize}{SIZE}
\SetKwFunction{FMax}{MAX}

\textbf{Require:} $I^{HR}$, $HR$, \FSize{$T^{HR}$}, $|\mathscr{N}|$\\
\SFn{\Fmain{}}{
\hspace{-9pt}$\triangleright$ \textbf{Step 0:} Get number of tiles\\
$|\mathscr{M}| \gets $\FSize{$I^{HR}$}/\FSize{$T^{HR}$}\\ 

\hspace{-9pt}$\triangleright$ \textbf{Step 1:} Encode all tiles using J2K\\
$\bigcup\limits_m T^{HR}_m.\text{\footnotesize\textit{J2K}}$ $\gets$ \FJcomp{$I^{HR}$, $\mathscr{M}$}\\
where $\mathscr{M}\coloneqq\{1,...,|\mathscr{M}|\}$\\

\hspace{-9pt}$\triangleright$ \textbf{Step 2:} Send tiles via channel\\
$\bigcup\limits_m T^{HR}_m.\text{\footnotesize\textit{J2K}}$ $\gets$ \FTR{$\bigcup\limits_m T^{HR}_m.\text{\footnotesize\textit{J2K}}$}\\

\hspace{-9pt}$\triangleright$ \textbf{Step 3:} Decode all tiles using J2K\\
$\bigcup\limits_m T^{HR}_m$ $\gets$ \FJdecomp{$\bigcup\limits_m T^{HR}_m.\text{\footnotesize\textit{J2K}}$, $\mathscr{M}$}\\

\hspace{-9pt}$\triangleright$ \textbf{Step 4:} Annotate high-resolution tiles using DL model\\
$BB_{DL}$, $CS_{DL}$ $\gets$ \FDL{$\bigcup\limits_m T^{HR}_m$, $\mathscr{M}$, $HR$}\\

\hspace{-9pt}$\triangleright$ \textbf{Step 5:} Spot indices of tiles that need human annotation\\
$\mathscr{N}$ $\gets$ \FIndex{$BB_{DL}$, $CS_{DL}$, $|\mathscr{N}|$}\\

\hspace{-9pt}$\triangleright$ \textbf{Step 6:} Extract chosen tiles from J2K\\
$\bigcup\limits_n T^{HR}_n.\text{\footnotesize\textit{J2K}}$ $\gets$
\FExt{$\bigcup\limits_m T^{HR}_m.\text{\footnotesize\textit{J2K}}$, $\mathscr{N}$, $HR$}\\

\hspace{-9pt}$\triangleright$ \textbf{Step 7:} Decode chosen high-resolution tiles using J2K\\
$\bigcup\limits_n T^{HR}_n$ $\gets$ \FJdecomp{$\bigcup\limits_n T^{HR}_n.\text{\footnotesize\textit{J2K}}$, $\mathscr{N}$}\\

\hspace{-9pt}$\triangleright$ \textbf{Step 8:} Annotate chosen tiles by human annotator\\
$BB_{HUM}$, $CS_{HUM}$ $\gets$ \FHUM{$\bigcup\limits_n T^{HR}_n$, $\mathscr{N}$}\\

\hspace{-9pt}$\triangleright$ \textbf{Step 9:} Make decision based on hybrid annotations\\
\textit{Decision} $\gets$ $BB_{HUM}$, $CS_{HUM}$, $BB_{DL}$, $CS_{DL}$\\

\KwRet{Decision}
}
\end{algorithm}

\vspace{-10mm}

\subsection{Proposed streamlined hybrid annotation framework}

\vspace{-1mm}

\begin{figure*}[!t]
  \centering
  \includegraphics[width=0.91\linewidth]{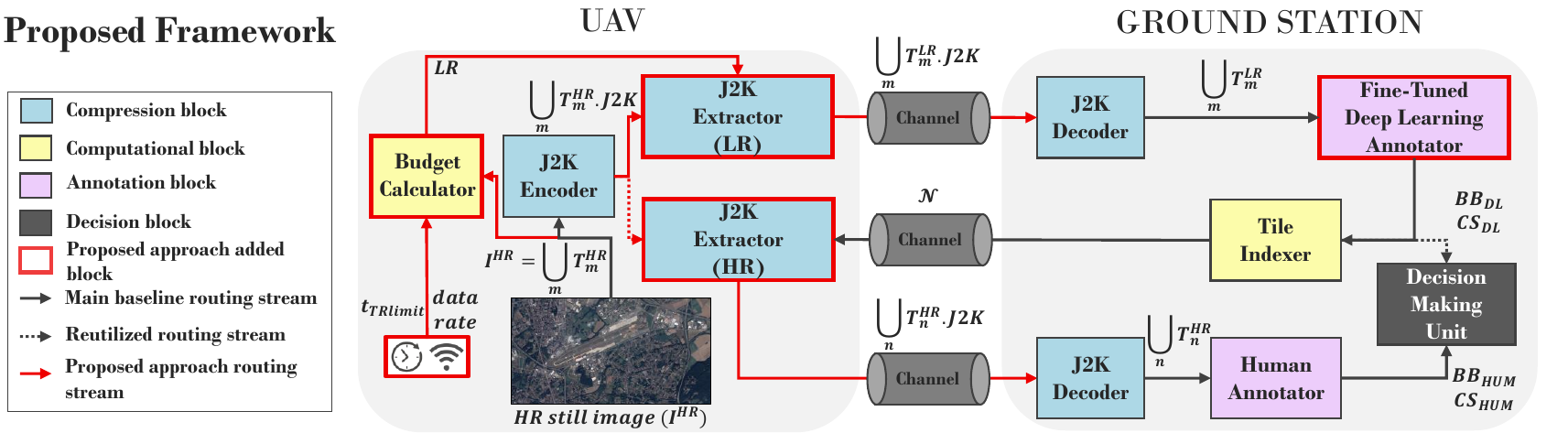} 
  \caption{Simplified block diagram of the proposed streamlined hybrid annotation framework.}
  \label{fig:proposed}
\end{figure*}

The objective of the proposed framework is to ensure that the decision-making process is rapid and reliable, meeting the critical needs of emergency response operations. To achieve this objective, we propose to modify the baseline framework by streamlining the hybrid annotation process to reduce the overall bandwidth consumption of the system. Figure \ref{fig:proposed} presents a simplified block diagram of the proposed streamlined hybrid annotation framework, highlighting the main differences with the baseline framework. A step-by-step pseudo-code of the proposed framework is presented in Algorithm 2. The streamlining process is achieved by implementing four improvements to the baseline framework: 

\newpage

\addtocounter{algocf}{-1}
\renewcommand{\algorithmcfname}{Subroutine}

\begin{algorithm}[!h]
\small
\caption{Indexer}
\DontPrintSemicolon
\SetKwProg{SFn}{Function}{:}{}
\SetKwFunction{FIndex}{IND}
\SetKwFunction{FMax}{MAX}

\SFn{\FIndex{$BB$, $CS$, $|\mathscr{Y}|$}}{
Sort $BB$ w.r.t. $CS$ values in ascending order\;
$\mathscr{X}, i \gets  \emptyset$ , 0\;
\While{$i < |\mathscr{Y}|$}{

    $\mathscr{X}.\text{append}(BB[i])$\;
    $i \gets i + 1$\;
}
$\mathscr{Y} \gets \mathscr{X}$\;
\KwRet $\mathscr{Y}$\;
}
\end{algorithm}

\vspace{-7.5mm}

\begin{algorithm}[!h]
\small
\caption{Extractor}
\DontPrintSemicolon  
\SetKwProg{SFn}{Function}{:}{}
\SetKwFunction{FExt}{EXT}
\SFn{\FExt{$\bigcup\limits_x T^{HR}_x.\text{\footnotesize\textit{J2K}}$, $\mathscr{Y}$, $R$}}{
\ForEach{$y \in \mathscr{Y}$}{
    Extract $T^R_y$ from  $\bigcup\limits_x T^{HR}_x.\text{\footnotesize\textit{J2K}}$ at resolution $R$
}
\KwRet $\bigcup\limits_y T^R_y.\text{\footnotesize\textit{J2K}}$
}
\end{algorithm}

\vspace{-5mm}

\begin{enumerate}[leftmargin=*]
    \item We define a \textit{budget calculator} block, detailed in Subroutine 3, that takes into account scenario-specific inputs, the transmission time limit ($t_{TRlimit}$) and available data rate ($data\_rate$) from the UAV, as well as the predetermined resolution levels ($Rlvls$) of the J2K encoding. The \textit{budget calculator} block then determines the highest resolution available ($LR$) at which we can send our images and sets the human annotation budget ($|\mathscr{N}|$) with the remaining bandwidth budget.    
    \item We fine-tune a DL model on low-resolution images to maintain optimal detection performance at the initial annotation step ($BB_{DL}$, $CS_{DL}$).
    \item We exploit the J2K scalable codestream by utilizing the tile extraction process at the UAV level, instead of at the ground station level, for both low-resolution and high-resolution tile extraction. 
    \item We reroute the framework to ensure that only the tiles selected for human annotation are sent in high-resolution, effectively reducing the overall bandwidth consumption of the framework.
\end{enumerate}

\newpage

\addtocounter{algocf}{-1}
\renewcommand{\algorithmcfname}{Algorithm}

\begin{algorithm}[!h]
\small
\caption{Proposed streamlined framework}
\DontPrintSemicolon
\SetKwProg{SFn}{Function}{:}{}
\SetKwFunction{Fmain}{main}
\SetKwFunction{Ftile}{BC}
\SetKwFunction{FJcomp}{J2K}
\SetKwFunction{FExt}{EXT}
\SetKwFunction{FJdecomp}{J2K$^{-1}$}
\SetKwFunction{FDL}{DL}
\SetKwFunction{FIndex}{IND}
\SetKwFunction{FHUM}{HUM}
\SetKwFunction{FTR}{TR}
\SetKwFunction{FSize}{SIZE}
\SetKwFunction{FMax}{MAX}

\textbf{Require: }$I^{HR}$, \FSize{$T^{HR}$}, $data\_rate$, $t_{TRlimit}$, $Rlvls$\\
\SFn{\Fmain{}}{
\hspace{-9pt}$\triangleright$ \textbf{Step 0:} Get number of tiles and chosen low resolution\\
$LR$, $HR$, $|\mathscr{N}|$, $|\mathscr{M}|$ $\gets$ \Ftile{$I^{HR}$, \FSize{$T^{HR}$}, $data\_rate$, $t_{TRlimit}$, $Rlvls$}\\

\hspace{-9pt}$\triangleright$ \textbf{Step 1:} Encode all tiles using J2K\\
$\bigcup\limits_m T^{HR}_m.\text{\footnotesize\textit{J2K}}$ $\gets$ \FJcomp{$I^{HR}$, $\mathscr{M}$}\\
where $\mathscr{M}\coloneqq\{1,...,|\mathscr{M}|\}$\\

\hspace{-9pt}$\triangleright$ \textbf{Step 2:} Extract all tiles in low resolution from J2K\\
$\bigcup\limits_m T^{LR}_m.\text{\footnotesize\textit{J2K}}$ $\gets$ \FExt{$\bigcup\limits_m T^{HR}_m.\text{\footnotesize\textit{J2K}}$, $\mathscr{M}$, $LR$}\\

\hspace{-9pt}$\triangleright$ \textbf{Step 3:} Send low-resolution tiles via channel\\
$\bigcup\limits_m T^{LR}_m.\text{\footnotesize\textit{J2K}}$ $\gets$ \FTR{$\bigcup\limits_m T^{LR}_m.\text{\footnotesize\textit{J2K}}$}\\

\hspace{-9pt}$\triangleright$ \textbf{Step 4:} Decode all low-resolution tiles using J2K\\
$\bigcup\limits_m T^{LR}_m$ $\gets$ \FJdecomp{$\bigcup\limits_m T^{LR}_m.\text{\footnotesize\textit{J2K}}$, $\mathscr{M}$}\\

\hspace{-9pt}$\triangleright$ \textbf{Step 5:} Annotate low-resolution tiles using DL model\\
$BB_{DL}$, $CS_{DL}$ $\gets$ \FDL{$\bigcup\limits_m T^{LR}_m$, $\mathscr{M}, LR$}\\

\hspace{-9pt}$\triangleright$ \textbf{Step 6:} Spot indices of tiles that need human annotation\\
$\mathscr{N}$ $\gets$ \FIndex{$BB_{DL}$, $CS_{DL}$, $|\mathscr{N}|$}\\

\hspace{-9pt}$\triangleright$ \textbf{Step 7:} Send chosen tile indices via channel\\
$\mathscr{N}$ $\gets$  \FTR{$\mathscr{N}$}\\

\hspace{-9pt}$\triangleright$ \textbf{Step 8:} Extract chosen tiles in high resolution from J2K\\
$\bigcup\limits_n T^{HR}_n.\text{\footnotesize\textit{J2K}}$ $\gets$
\FExt{$\bigcup\limits_m T^{HR}_m.\text{\footnotesize\textit{J2K}}$, $\mathscr{N}$, $HR$}\\

\hspace{-9pt}$\triangleright$ \textbf{Step 9:} Send chosen high-resolution tiles via channel\\
$\bigcup\limits_n T^{HR}_n.\text{\footnotesize\textit{J2K}}$ $\gets$  \FTR{$\bigcup\limits_n T^{HR}_n.\text{\footnotesize\textit{J2K}}$}\\

\hspace{-9pt}$\triangleright$ \textbf{Step 10:} Decode chosen high-resolution tiles using J2K\\
$\bigcup\limits_n T^{HR}_n$ $\gets$ \FJdecomp{$\bigcup\limits_n T^{HR}_n.\text{\footnotesize\textit{J2K}}$, $\mathscr{N}$}\\

\hspace{-9pt}$\triangleright$ \textbf{Step 11:} Annotate chosen tiles by human annotator\\
$BB_{HUM}$, $CS_{HUM}$ $\gets$ \FHUM{$\bigcup\limits_n T^{HR}_n$, $\mathscr{N}$}\\

\hspace{-9pt}$\triangleright$ \textbf{Step 12:} Make decision based on hybrid annotations\\
\textit{Decision} $\gets$ $BB_{HUM}$, $CS_{HUM}$, $BB_{DL}$, $CS_{DL}$\\

\KwRet{Decision}
}
\end{algorithm}

\vspace{-5mm}

In summary, the proposed framework utilizes J2K's scalable codestream to reduce the overall bandwidth consumption by first sending a low-resolution image for initial DL network annotation and then transmitting only the selected high-resolution tiles for human expert annotation.

\newpage

\addtocounter{algocf}{0}
\renewcommand{\algorithmcfname}{Subroutine}

\begin{algorithm}[!h]
\small
\caption{Budget calculator}
\DontPrintSemicolon  
\SetKw{KwTo}{to} 
\SetKw{KwBreak}{break} 
\SetKwProg{SFn}{Function}{:}{}
\SetKwFunction{Ftile}{BC}
\SetKwFunction{FSize}{SIZE}
\SetKwFunction{FMax}{MAX}

\SFn{\Ftile{$I^{HR}$, \FSize{$T^{HR}$}, $data\_rate$, $t_{TRlimit}$, $Rlvls$}}{
    $LR \gets -1$\;
    $|\mathscr{N}|, |\mathscr{M}| \gets 0$\;
    $BW \gets data\_rate \cdot t_{TRlimit}$ \;
    $HR \gets $\FMax{$Rlvls$}\;
    \For{$R \gets HR$ \KwTo $1$}{
        \If{\FSize(${I^R}) \leq BW$}{
            $LR \gets R$ \;
            $|\mathscr{N}| \gets (BW - $\FSize{$I^R$}) / \FSize{$T^{HR}$}\;
            \KwBreak
        }
    }
    \If{$LR = -1$}{
        \textbf{Insufficient BW Budget, must relax constraints}\;
    
    }
    $|\mathscr{M}| \gets $\FSize{$I^{HR}$}/\FSize{$T^{HR}$}\;

    \KwRet{$LR$, $HR$, $|\mathscr{N}|$, $|\mathscr{M}|$}

}
    \end{algorithm}

\vspace{-6mm}

\section{Experimental setup}
\label{sec:implementation}
In this section, we introduce the materials used, the metrics applied and the experiments done to evaluate the proposed framework performance.

\vspace{-3mm}

\subsection{Materials}

\vspace{-2mm}

The dataset used is composed of 136 high-resolution satellite still images representing airports in different locations \cite{Palmaerts_2023}. We train a YOLOv8\footnote{https://github.com/ultralytics/ultralytics} model for multi-class identification of civilian aircraft.
Five models are fine-tuned on five resolution levels with a training/validation/test sets split of 89/42/5 images, respectively. The J2K encoding, extraction and decoding tasks, are performed using the OpenJPEG platform\footnote{https://www.openjpeg.org/}.

\subsection{Evaluation metrics}
Two evaluation metrics are used to compare the baseline and the proposed frameworks: \textbf{response time ratio} (${t_{RS}}\_ratio$) and \textbf{recall difference} ($recall\_diff$).\\

${t_{RS}}\_ratio$ is used to compare the response times between the proposed and the baseline frameworks. For each framework, the response time ($t_{RS}$), the time at which the decision is taken, is defined as:
\begin{equation}
t_{RS} = t_{TR} + t_{HUM}
\end{equation}
where $t_{TR}$ is the transmission time and $t_{HUM}$ is the human annotator time. \\

\noindent The transmission time ($t_{TR}$) is the time needed for data to be transmitted through the channel. We distinguish between the transmission times of the two frameworks as follows. \\

\noindent The baseline framework transmission time is defined as:
\begin{equation}
    t_{TR} = \frac{\scriptsize{\FSize(\bigcup\limits_m T^{HR}_m.\text{\footnotesize\textit{J2K}})}}{\scriptsize{data\_rate}}
\end{equation}
where $data\_rate$ varies depending on the scenario at hand.\\

\noindent The proposed framework transmission time is defined as: 
\begin{equation}
    t_{TR} = \frac{\scriptsize{\FSize}\scriptsize{(\bigcup\limits_m T^{LR}_m.\textit{J2K}) + \scriptsize{\FSize}\scriptsize{(\bigcup\limits_n T^{HR}_n.\textit{J2K})}}}{\scriptsize{data\_rate}}
\end{equation}
where the framework adapts the tile resolution ($LR$) depending on the $data\_rate$ and transmission time limit $t_{TRlimit}$ of the scenario at hand.\\ 

\noindent It should be noted that, for both frameworks, we consider that any additional transmitted metadata is minor in size compared to the image data and therefore has a negligible impact on the total data size.\\

\noindent The human annotator time ($t_{HUM}$), the time required for human experts to annotate uncertain areas of an image, is defined as:
\begin{equation}
t_{HUM} = \mu_{t_{\text{HUM}}} \cdot |\mathscr{N}|
\end{equation}
where $\mu_{t_{\text{HUM}}}$ is the average time a human expert needs to annotate a single tile and $|\mathscr{N}|$ is the number of tiles requiring expert annotation.\\

\noindent The \textbf{response time ratio} (${t_{RS}}\_ratio$) is thus defined as:
\begin{equation}
    {t_{RS}}\_ratio = \frac{t_{RS}(Baseline)}{t_{RS}(Proposed)}
\end{equation}

\vspace{4.5mm}

$recall\_diff$ quantifies the difference in object detection performance between the proposed and the baseline frameworks. We consider the $recall$ as the metric for detection accuracy because, in an emergency situation, the priority is to ensure that we have as few false negatives as possible to avoid missing objects of interest.\\

\noindent The $recall$ is defined as: 
\begin{equation}
    recall = \frac{TP}{TP+FN}
\end{equation}
where $TP$ is the number of true positive and $FN$ is the number of false negative, a detection being considered as a $TP$ if the intersection over union with the ground truth is higher than 0.1 to prevent missing true positive at low resolution, objects being represented by only few pixels. \\

\noindent The \textbf{recall difference} ($recall\_diff$) is thus defined as:

\vspace{-4mm}

\begin{align}
    recall\_diff &= recall(t_{RS}(Baseline)) \nonumber \\
    &\quad - recall(t_{RS}(Proposed))
\end{align}

\subsection{Experiments}

The objective of the proposed framework is to decrease the response time in emergency scenarios compared to a conventional baseline approach while maintaining optimal object detection. 
The use case chosen is a surveillance scenario dedicated to the identification of civilian aircraft in multiple airport settings.  We configure the J2K compression with five resolution levels and also fine-tune the YOLOv8 model on the same five resolution levels. We test the frameworks on high-resolution images of 60 MP and examine nine emergency scenarios whilst varying two scenario-specific parameters: $t_{TRlimit}$ dependent on the urgency of the emergency response, and the available $data\_rate$. We fix three $t_{TRlimit}$ values of 3, 10, and 30 minutes to emulate three levels of urgency in emergency response scenarios. We consider three $data\_rate$ values: 22, 88, and 176 kbps. These rates fall within the minimum-to-median range available for the Iridium Certus satellite network, the standard for low-bandwidth UAV communications\footnote{https://www.iridium.com/iridiumcertus/}. 
We do not consider the high data rate range offered by the Iridium Certus network, as we would obtain equivalent performances between our framework and the baseline. We fix $\mu_{t_{\text{HUM}}}$ at 0.5 minutes/tile.

\begin{figure*}[!t]
  \centering
  \includegraphics[width=0.98\linewidth]{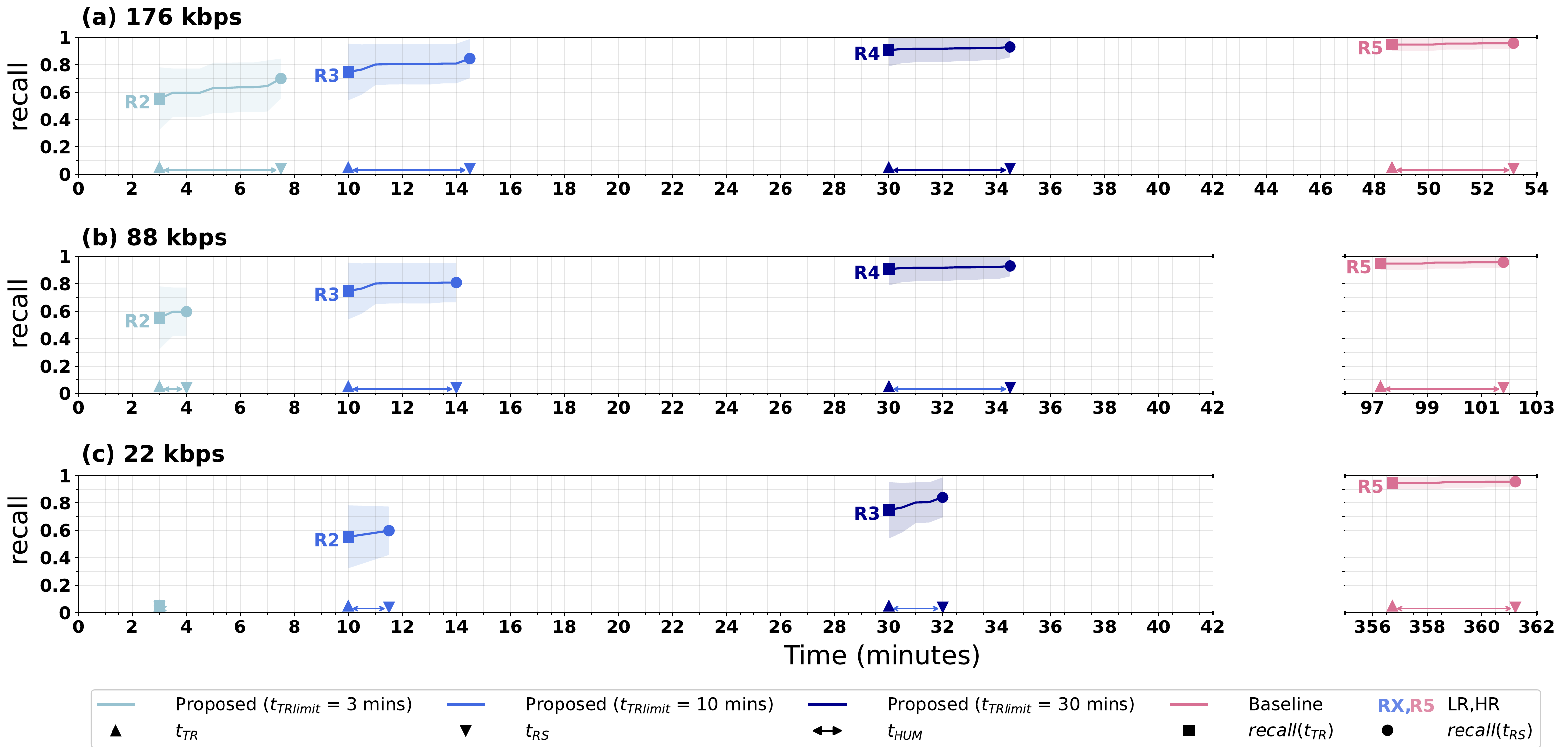} 
  \caption{Recall versus time plot for the baseline framework and the proposed framework for high-resolution image size of 60 MP under a data rate limit of (a) 176 kbps (b) 88 kbps and (c) 22 kbps, with a maximum human annotator time of 5 minutes and considering emergency scenarios with a transmission time limit of: 3, 10, and 30 minutes.}
  \label{fig:recall}
\end{figure*}

\section{Results and discussion}

The comparison of the proposed and the baseline frameworks for the nine scenarios previously mentioned can be seen in Figure \ref{fig:recall}. We start by analyzing the least constrained case with a $data\_rate$ of 176 kbps (Figure \ref{fig:recall}a). When considering $t_{TRlimit}$ values of 3, 10 and 30 minutes, we respectively obtain a penalizing $recall\_diff$ of 0.257, 0.112 and 0.028 while observing a gain in $t_{RS}\_ratio$ of 7.08, 3.67 and 1.54. Indeed, there is a trade-off between the $t_{RS}\_ratio$ and the $recall\_diff$ when comparing the baseline and the proposed approaches. This trade-off is a direct consequence of the proposed framework's streamlining approach. Given that the proposed framework takes into consideration the scenario-specific bandwidth constraints, it prioritizes the transmission of lower-resolution images for initial annotation. This explains the large improvement in $t_{RS}\_ratio$ compared to the baseline approach albeit with a slight penalty in $recall\_diff$. On the other hand, given that the baseline framework does not employ a streamlining strategy, it ends up always sending the still image at the highest possible resolution. This explains its slight improvement in $recall\_diff$, however this comes with a large penalty in $t_{RS}\_ratio$. The same observations can be made at a $data\_rate$ of 88 kbps (Figure \ref{fig:recall}b). We respectively observe for the three $t_{TRlimit}$ values of 3, 10 and 30 minutes that the $recall\_diff$ worsens to 0.35, 0.148 and 0.028 but the $t_{RS}\_ratio$ improves to 25.44, 7.27 and 2.95. The streamlining strategy of the proposed framework explains its robustness against more constraint scenarios, allowing to send the same low resolution for the initial annotation step. The only downside being the reduction of $t_{HUM}$ due to the lower bandwidth budget, explaining the slight reduction of $recall(t_{RS})$. In contrast, the baseline framework's $t_{RS}$ performance deteriorates given that it does not adapt to the scenario-specific constraints. Figure \ref{fig:recall}c presents the scenarios at the smallest $data\_rate$ of 22 kbps. This constraint tests the limits of the proposed framework. For the scenario with a $t_{TRlimit}$ of 3 minutes, the bandwidth constraint is too restrictive, even the lowest resolution cannot be sent, leading to a $recall(t_{RS})$ of 0. For the scenarios with a $t_{TRlimit}$ of 10 and 30 minutes, we obtain a $recall\_diff$ of 0.359 and 0.115 respectively. However, we observe staggering gain in $t_{RS}$ of 34.4 and 11.28 respectively. These very-low bandwidth scenarios clearly show the gain in $t_{RS}$ of the proposed streamlined framework compared to the baseline framework. 


\section{Conclusion}
\label{sec:Conc}

\vspace{0.75mm}

We present a streamlined hybrid annotation framework to enhance UAV object detection in bandwidth-restricted scenarios. The framework successfully combines automated DL-based detection with human expertise to refine annotations in uncertain areas, managing to cut the decision-making time by up to 34 times for emergency response operations. Future works will focus on optimizing each of the subroutines within the proposed framework, primarily focusing on improving both the fine-tuned DL model and the tile selection strategy for human annotation. 
Additionally, employing a few-shot learning strategy can improve model adaptability in emergency scenarios. Furthermore, implementing an active learning layer can benefit the model's long-term training.

\vspace{0.75mm}

\section{Acknowledgments}
\label{sec:ACK}

\vspace{0.75mm}

The Pléiades images were downloaded from the \textit{Pléiades 4 Belgium} platform, through an agreement with Belgian Science Policy Office (\textit{BELSPO}). 
The images were labeled by the Institut Scientifique de Service Public (\textit{ISSeP}).
Tiffanie Godelaine is supported by \textit{MedReSyst}, funded by the Walloon Region and European Union \textit{Wallonie 2021-2027} program.

\newpage


\end{document}